\documentstyle[prb,twocolumn,aps,graphicx]{revtex}
\tighten 

\begin{document}
\draft

\wideabs{
\title{A prohibition of equilibrium spin currents in multi-terminal ballistic devices}
\author{A. A. Kiselev\cite{email} and K. W. Kim}
\address{Department of Electrical and Computer Engineering,
North Carolina State University, Raleigh, NC 27695-7911}

\maketitle
\begin{abstract}
We show that in the multi-terminal ballistic devices with
intrinsic spin-orbit interaction connected to normal metal
contacts there are no equilibrium spin currents present at any
given electron energy. Obviously, this statement holds also after
the integration over all occupied states. Based on the proof of
this fact, a number of scenarios involving nonequilibrium spin
currents is identified and further analyzed. In particular, it is
shown that an arbitrary two-terminal device cannot polarize
transient current. The same is true for the output terminal of an
$N$-terminal device when all $N-1$ inputs are connected in
parallel.
\end{abstract}

}

%%%%%%%%%%%%%%%%%%%%%%%%%%%%%%%%%%%%%%%%%%%%%%%%%%%%%%%%%%%%%%%%%%%%%%%
%%%%%%%%%%%%%%%%%%%%%%%%%%%%%%%%%%%%%%%%%%%%%%%%%%%%%%%%%%%%%%%%%%%%%%%
%\section{Introduction}
%\label{introduction}

The results of Murakami {\it et al.}\cite{Murakami} and Sinova
{\it et al.}\cite{Sinova} excited spintronic community by
suggesting presence of the dissipationless spin currents in
electron/hole systems with intrinsic spin-orbit (SO) interaction.
A quick and massive response by quite a few of the research teams
approached varied levels of generality and with mixed
conclusions.\cite{dissipationless}

The matter was further highlighted by Rashba\cite{Rashba2003} who
distilled the problem of \emph{equilibrium} dissipationless spin
currents in a two-dimensional (2D) electron gas into an unobscured
puristic form of a paradox.

A similar phenomenology could potentially exist in 1D systems,
especially in relation to the complex multi-terminal ballistic
devices. Rather obvious arguments (see, \emph{e.g.},
Ref.~\onlinecite{Mal'shukov}), although, put a ban on the
existence of equilibrium spin currents in an ideal 1D wire with
parabolic dispersion and linear-in-$k$ SO coupling. Nevertheless,
for the 1D structures with static scatters, the opposite
conclusions were recently announced\cite{Pareek} (but, again,
their validity is now also questioned\cite{Nikolic}).

In this Report we provide an unambiguous proof of the fact that in
multi-terminal ballistic devices equilibrium spin currents do not
exist. This conclusion is universal and does not depend on the
particular choice of the electron Hamiltonian or the particular
type of the SO interaction.

%%%%%%%%%%%%%%%%%%%%%%%%%%%%%%%%%%%%%%%%%%%%%%%%%%%%%%%%%%%%%%%%%%%%%%%
%%%%%%%%%%%%%%%%%%%%%%%%%%%%%%%%%%%%%%%%%%%%%%%%%%%%%%%%%%%%%%%%%%%%%%%
%\section{Theory}
%\subsection{Structure}

\begin{figure}
\begin{center}
\includegraphics*[bb=43 280 523 510,width=110mm,angle=270]{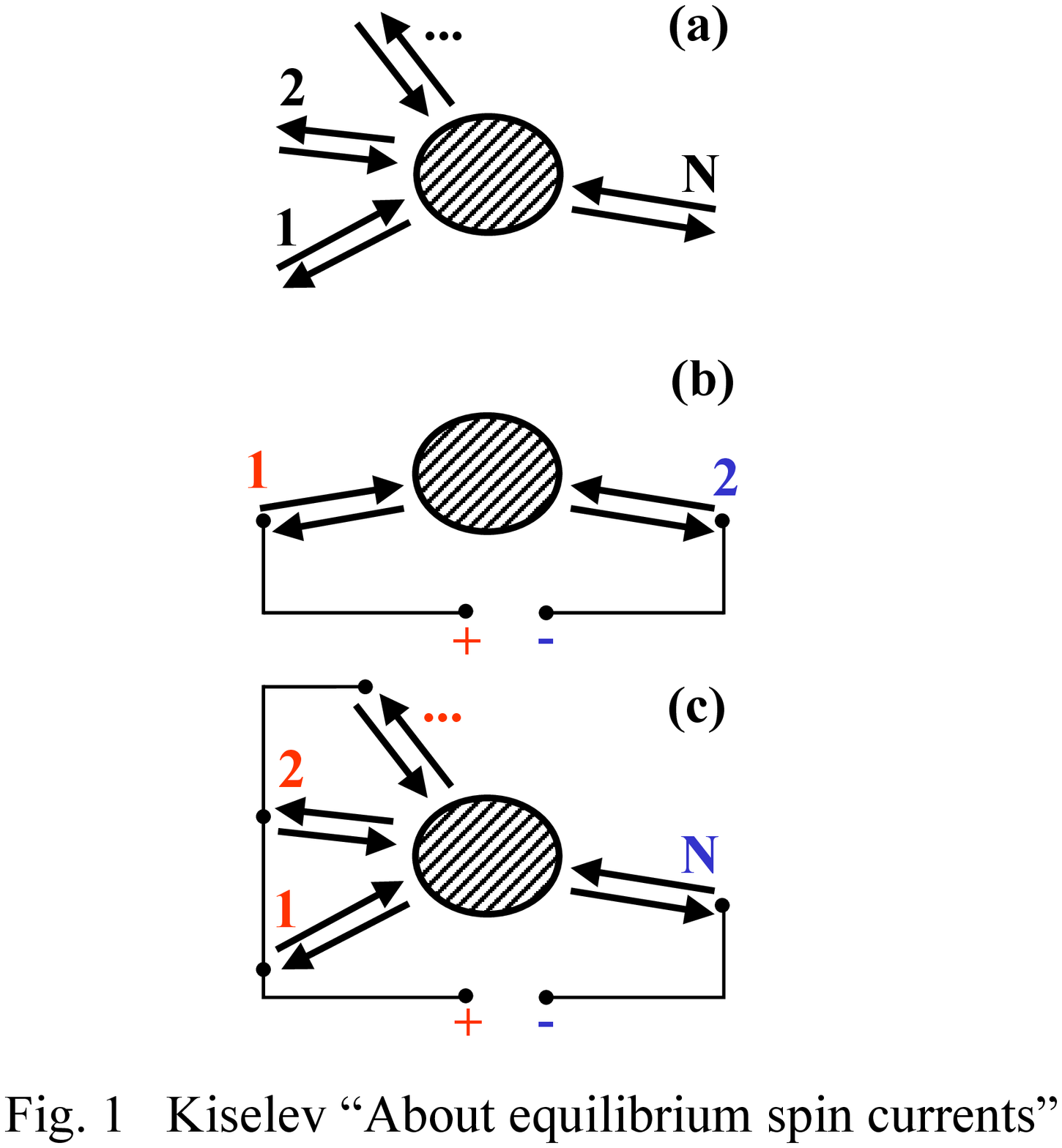}
\end{center}
\caption{\label{f_sketch} (a) $N$-terminal ballistic device with
intrinsic SO interaction (in shaded area). Arrows show direction
of input and output fluxes through connecting terminals; (b)
two-terminal device with bias applied; (c) $N$-terminal device
with $N-1$ inputs connected in parallel.}
\end{figure}

We consider a ballistic device consisting of a structure of
complex shape (for a simple example, some area, ``stamped out''
from the two-dimensional electron gas) that is connected to the
{\it exterior} by a number of quasi-1D wires (arms)---see
Fig.~\ref{f_sketch}~(a) for a visual guide.

%%%%%%%%%%%%%%%%%%%%%%%%%%%%%%%%%%%%%%%%%%%%%%%%%%%%%%%%%%%%%%%%%%%%%%%
%%%%%%%%%%%%%%%%%%%%%%%%%%%%%%%%%%%%%%%%%%%%%%%%%%%%%%%%%%%%%%%%%%%%%%%
%\subsection{Scattering matrix}

To analyze the static linear-response conductances of the
multi-terminal ballistic device we apply the Landauer-B\"uttiker
formalism,\cite{scattering_matrix} that we have extended to
incorporate also spin-related
phenomena.\cite{Kiselev_APL2001,Kiselev_JAP2003} Following
closely these earlier publications, we now briefly introduce
relevant basic concepts.

A finite-size \emph{scattering matrix} ${\bf
S}$,\cite{scattering_matrix}
\begin{equation}\label{scat_matrix}
{\bf S}=\left(
\begin{array}{cccc}
  r_{11} & t_{12} & \cdots & t_{1N} \\
  t_{21} & r_{22} & \cdots & t_{2N} \\
  \vdots & \vdots & \ddots & \vdots \\
  t_{N1} & t_{N2} & \cdots & r_{NN}
\end{array}
\right)
\end{equation}
describes in a condensed form transmission and reflection
properties of an arbitrarily complex linear (ballistic) system
connected to the exterior via 1D wires. It consists of the
(diagonal) reflection coefficients $r_{ii}$ (in the $i$-th
channel) and (off-diagonal) transmission coefficients $t_{ij}$
(describing propagation of the particle from the $j$-th into the
$i$-th channel, $i\neq j$). For every electron energy we choose
to enumerate as distinguishable channels (and terminals) all
energetically allowed electron fluxes through different 1D
subbands (transverse modes), even in the same wire. Thus, the
number of the channels can differ substantially from the number
of the attached contacts. Because of this fact, biases applied to
different terminals not always can be modulated independently.
Since an elementary electron flux $F$ carried by a particular
ideal 1D channel is defined by product of square of the electron
plane wave amplitude $\psi$ and the state group velocity
$v=\partial H/\partial k$, it is a usual practice to define the
scattering matrix in terms of the \emph{current amplitudes}
$u=\sqrt{v}\psi$. Thus, $F\propto u^+ u$.

For a spinless particle, the coefficients $r_{ii}$, $t_{ij}$ are
scalars. By taking electron spin into consideration, the number of
transmission channels effectively doubles. For this discussion, we
will assume that in the vicinity of the normal metal contacts the
channel asymmetry is either not present or effectively
\emph{screened} by the metal. Thus, the spin-dependent
interactions are spatially limited to the {\it interior} of the
system, and the transport in the leads through two degenerate spin
subchannels is coherent and requires some cautious treatment,
especially when evaluating total channel fluxes and their
polarization.

To incorporate into our formalism this spin-related
channel-doubling, we convert transmission and reflection
coefficients of the ${\bf S}$ into $2\times2$ submatrices. Each
$2\times2$ submatrix $\hat{x}$ ($\hat{x}=\hat{r}_{ii}$ or
$\hat{t}_{ij}$) in this case can be equivalently expanded as
\begin{equation}\label{expansion}
\hat{x}=\left(
\begin{array}{cc}
x_{\uparrow\uparrow}&x_{\uparrow\downarrow}\\
x_{\downarrow\uparrow}&x_{\downarrow\downarrow}\\
\end{array}
\right)=\hat{1}x_1+i\sum_\alpha \hat{\sigma}_\alpha x_\alpha\:,
\end{equation}
with a small additional benefit of conventional algebraic
manipulations with matrix objects. Here $\hat{1}$ is a unit
$2\times2$ matrix, $\hat{\sigma}_\alpha$($\alpha=x,y,z$) are the
usual Pauli matrices ($\uparrow$~and $\downarrow$ utilized here
are eigenstates of the operator $\hat{\sigma}_z$, with axis $z$
actually arbitrarily chosen).

In case of degenerate spin subchannels, an elementary channel
flux $F\propto\hat{u}^+\hat{u}$ (input or output) is conveniently
given by the spinor column $\hat{u}=(u_\uparrow,u_\downarrow)$.

%%%%%%%%%%%%%%%%%%%%%%%%%%%%%%%%%%%%%%%%%%%%%%%%%%%%%%%%%%%%%%%%%%%%%%%
%%%%%%%%%%%%%%%%%%%%%%%%%%%%%%%%%%%%%%%%%%%%%%%%%%%%%%%%%%%%%%%%%%%%%%%
%\subsection{Partially polarized flux}

The relative magnitude and phase between $u_\uparrow$ and
$u_\downarrow$ characterize orientation of the spin; vector
$\bbox{P}$, defined by three components $P_\alpha$ in the $xyz$
coordinate system, is called the polarization vector, $P_\alpha
F\propto\hat{u}^+\hat{\sigma}_\alpha\hat{u}$. For the arbitrary
spinor $\hat{u}$ the absolute value $|\bbox{P}|=1$. When the
incident flux $\hat{u}$ is pumped into the channel $j$, the
$\hat{t}_{ij}\hat{u}$ part of it will seep through the structure
into the $i$-th output channel. In the general case of
spin-dependent interactions present in the system,
$\hat{t}_{ij,\alpha}$ are potentially non-zero and the spin will
rotate from its original orientation. Moreover, the {\it
magnitude} of the transmitted flux can, in principle, now depend
on the spin orientation of the incident flux.

\emph{Partially polarized} electron fluxes can be mimicked  by a
number of independent (not phase coherent) elementary fluxes
$\hat{u}_q$ through the same channel that are just additive in
the case of our linear system
\begin{equation}\label{sum_of_fluxes}
F=\sum_q F_q,\ \bbox{P} F=\sum_q (\bbox{P}F)_q.
\end{equation}
To present the unpolarized input flux in channel $j$, one can use,
for example, two elementary fluxes $\hat{u}_1=(1,0)$ and
$\hat{u}_2=(0,1)$. Now it is very easy to evaluate $F$ and
$\bbox{P}$ for the electron flux, transmitted into channel $i$.
Indeed,
\begin{equation}\label{total}
F\propto|t_{ij,1}|^2+\sum_\alpha |t_{ij,\alpha}|^2,
\end{equation}
with the $\alpha$-component of spin polarization
\begin{equation}\label{polarization}
P_\alpha F\propto2{\rm
Im}(t_{ij,1}t_{ij,\alpha}^*+t_{ij,\alpha+1}^*t_{ij,\alpha+2}),
\end{equation}
where $xyz$ indices are cyclically permuted.\cite{one_component}

%%%%%%%%%%%%%%%%%%%%%%%%%%%%%%%%%%%%%%%%%%%%%%%%%%%%%%%%%%%%%%%%%%%%%%%
%%%%%%%%%%%%%%%%%%%%%%%%%%%%%%%%%%%%%%%%%%%%%%%%%%%%%%%%%%%%%%%%%%%%%%%
%\subsection{Symmetry analysis}

There are strict fundamental limitations on the components of the
scattering matrices. Hereafter we present an adaptation of these
limitations suitable for the devices with the degenerate spin
subchannels in the leads.\cite{Kiselev_JAP2003}

With the total flux ${\bf F}\propto {\bf U}^+{\bf U}$, the
requirement of flux conservation for ${\bf U}_{\rm out}={\bf
S}{\bf U}_{\rm in}$ corresponding to an arbitrary column ${\bf
U}_{\rm in}$ (symbolically representing coherent incident plane
waves $\hat{u}_i$ coming through all channels), can only be
guaranteed if
\begin{equation}\label{flux_conservation}
{\bf S}^+{\bf S}={\bf 1}.
\end{equation}

For a broad class of problems, including the one with intrinsic SO
coupling, time-reversal invariance (with the operator
$\hat{T}=-i\hat{\sigma}_yK$ where $K$ is the complex conjugation)
establishes the following relation on the scattering matrix ${\bf
S}$
\begin{equation}\label{time_inversion}
\hat{\sigma}_y{\bf S}^*\hat{\sigma}_y{\bf S}={\bf 1}.
\end{equation}
This equation should be regarded as a symbolic one, with the Pauli
matrix multiplications applied to each submatrix $\hat{r}_{ii}$,
$\hat{t}_{ij}$ separately.

Combined with Eq.~(\ref{flux_conservation}), this relation can be
converted into a more practical form $\hat{\sigma}_y{\bf
S}^*\hat{\sigma}_y={\bf S}^+$, that immediately results in
\begin{equation}\label{r&t_components}
r_{ii,\alpha}=0, \ t_{ij,1}= t_{ji,1},\
t_{ij,\alpha}=-t_{ji,\alpha},
\end{equation}
Thus, the polarization of the {\it back-reflected} flux from
unpolarized source is not possible.\cite{no_SO}

Additional structure symmetry elements, when present, could
provide further restricting relations on the components of ${\bf
S}$ (see Ref.~\onlinecite{Kiselev_JAP2003} for details). However,
they are not necessary for the further consideration.

%%%%%%%%%%%%%%%%%%%%%%%%%%%%%%%%%%%%%%%%%%%%%%%%%%%%%%%%%%%%%%%%%%%%%%%
%%%%%%%%%%%%%%%%%%%%%%%%%%%%%%%%%%%%%%%%%%%%%%%%%%%%%%%%%%%%%%%%%%%%%%%
%\section{Theorem}

With all relevant basic quantities and concepts already
introduced, it is now time to formulate and prove the following
{\bf theorem}: \emph{For an arbitrary multiterminal ballistic
device with intrinsic SO interaction that is connected to the
normal metal contacts the equilibrium electron flux going out of
the device through any particular terminal is unpolarized,
i.e.}\cite{i_neq_j}
\begin{equation}\label{theorem}
\sum_j (\bbox{P}F)_{ij}\equiv 0.
\end{equation}

We start the \emph{proof} of this theorem with
Eq.~(\ref{flux_conservation}) and write out explicitly the
$ii$-th matrix element of this equation
\begin{equation}\label{ii_term}
\sum_j (\bbox{S}^+)_{ij}(\bbox{S})_{ji}=\hat{1}\:,
\end{equation}
with
\[
(\bbox{S}^+)_{ij}=\hat{t}_{ji}^+=\hat{1}t_{ji,1}^*-i\sum_\alpha
\hat{\sigma}_\alpha t_{ji,\alpha}^*\:.
\]
After performing summation over $i$ explicitly and collecting
terms with $\hat{\sigma}_\alpha$ on the \emph{l.h.s.} of the
equation (no such terms present on the \emph{r.h.s.}) we get
\begin{eqnarray}\label{sum_explicit}
...+i\hat{\sigma}_\alpha&\sum_{j\neq i} (&t_{ji,1}^*t_{ji,\alpha}
-t_{ji,\alpha}^*t_{ji,1}\\
&&\mbox{}+t_{ji,\alpha+1}^*t_{ji,\alpha+2}
-t_{ji,\alpha+2}^*t_{ji,\alpha+1})+...\nonumber
\end{eqnarray}
Making use of Eq.~(\ref{r&t_components}), the equation under the
sign of summation is exactly $2{\rm
Im}(t_{ij,1}t_{ij,\alpha}^*+t_{ij,\alpha+1}^*t_{ij,\alpha+2})$
which, according to Eq.~(\ref{polarization}) is an
$\alpha$-polarized part of the flux $(P_\alpha F)_{ij}$ going out
of the device through terminal $i$, induced by the
\emph{unpolarized} incident flux through terminal $j$ (and the
incident electron fluxes are naturally unpolarized in case of
normal metal terminals). Thus, it turns out that in order to
satisfy the flux conservation condition of
Eq.~(\ref{flux_conservation}), $\sum_j (\bbox{P}F)_{ij}$ should
be zero, that concludes the proof.

Since this statement holds for any incident electron energy
$E_{\text{kin}}$ independently, it holds, naturally, also after
integration over all occupied states.

%\section{Consequences}

When the infinitesimal biases are applied to terminals,
uncompensated {\it nonequilibrium} currents will flow through the
device. In presence of SO interactions, these currents can
generally become polarized (like, for example, it happens in the
case of the T-shaped spin filter\cite{Kiselev_APL2001}). However,
there are situations, when, as a consequence of the theorem
proven above, these polarization processes are \emph{fundamentally
prohibited.}
%\subsection{Two-terminal device}

(\emph{i}) \emph{Two-terminal device.}---An arbitrary system with
just two connecting terminals [Fig.~\ref{f_sketch}~{b}] cannot
polarize transmitted electron flux (this finding was earlier
announced in Ref.~\onlinecite{Kiselev_JAP2003}; see also
Ref.~\onlinecite{two_terminals} for the discussion of related
phenomena). Indeed, the sum in Eq.~(\ref{theorem}) reduces in this
case to a single term that is bound, by the theorem, to be zero.

%\subsection{Multiterminal device}
(\emph{ii}) \emph{$N$-terminal device with $N-1$ inputs connected
in parallel.}---When all but one terminal are kept at the same
bias [Fig.~\ref{f_sketch}~(c)], the output current through that
last terminal is unpolarized. Again, for the $N$-th terminal we
return exactly to the situation described by the theorem. For a
not-so-obvious example, let us consider some ballistic device
that is connected to the exterior by just two physical wires, one
of them supporting multi-channel transport and another one
allowing only a single propagating mode (apart from spin
degeneracy): current in this second wire is \emph{always}
unpolarized.

%%%%%%%%%%%%%%%%%%%%%%%%%%%%%%%%%%%%%%%%%%%%%%%%%%%%%%%%%%%%%%%%%%%%%%%
%%%%%%%%%%%%%%%%%%%%%%%%%%%%%%%%%%%%%%%%%%%%%%%%%%%%%%%%%%%%%%%%%%%%%%%
%\section{Summary}

\emph{In summary,} we have formulated and proven a theorem
relating very general and fundamental properties of a ballistic
system (current conservation and the symmetry in respect to the
time inversion) with the ability of this system to polarize
transient currents. When all input currents are supplied by
unpolarized sources, each and every output current is also
unpolarized for the system in equilibrium, \emph{i.e.}, no
equilibrium spin currents are allowed.

%\section{Acknowledgements}

{\it Acknowledgements.}---This work was supported, in part, by
DARPA, ONR, and SRC/MARCO.

%%%%%%%%%%%%%%%%%%%%%%%%%%%%%%%%%%%%%%%%%%%%%%%%%%%%%%%%%%%%%%%%%%%%%%%
%%%%%%%%%%%%%%%%%%%%%%%%%%%%%%%%%%%%%%%%%%%%%%%%%%%%%%%%%%%%%%%%%%%%%%%

%\vspace*{-0.6cm}

%%%%%%%%%%%%%%%%%%%%%%%%%%%%%%%%%%%%%%%%%%%%%%%%%%%%%%%%%%%%%%%%%%%%%%%
%%%%%%%%%%%%%%%%%%%%%%%%%%%%%%%%%%%%%%%%%%%%%%%%%%%%%%%%%%%%%%%%%%%%%%%

\begin{references}
\bibitem{one_component} This equation shows, in particular,
that if only one component of $\hat{t}_{ij}$ is present,
polarization of the transmitted electron flux is zero for
unpolarized input.
\bibitem{no_SO} Note, that
without the SO term the operator $K$ itself and any combination
$\hat{\sigma}_\alpha K$ would commute with the Hamiltonian, so
one would obtain $t_{ij,\alpha}\equiv0$. This would indicate that
there are no spin-polarizing effects in {\it transmission} as
well, which is in agreement with common sense.
\bibitem{i_neq_j} Because the back-scattered flux is always
unpolarized, it does not matter whether the $i=j$ term is
included into the summation or not.
\end{references}
\end{document}